\documentclass{ws-ijmpa}

\begin{document}

\markboth{Chad Finley}
{Angular Correlation Estimates for Ultrahigh Energy Cosmic Rays}

%
\catchline{}{}{}{}{}
%

\title{Angular Correlation Estimates for Ultrahigh Energy Cosmic Rays}

\author{\footnotesize CHAD FINLEY\footnote{finley@phys.columbia.edu}\\ 
for the High Resolution Fly's Eye (HiRes) 
Collaboration\footnote{see http://hires.phys.columbia.edu 
for a complete list of authors}}

\address{Physics Department, Columbia University,\\ 538 West 120th Street,
New York, NY 10027, USA}

\maketitle

\begin{abstract}
Anisotropy in arrival directions of ultrahigh energy cosmic rays
offers the most direct way to search for the sources of these particles.
We present estimates
of the angular correlation in the HiRes sample of stereo events
above $10^{19}$\,eV, and in the combined sample of HiRes and AGASA 
events above $4\times 10^{19}$\,eV.

\keywords{ultrahigh energy cosmic rays; anisotropy of cosmic rays}
\end{abstract}

\section{Introduction}

Identifying the sources of ultrahigh energy cosmic rays remains
one of the major challenges in astrophysics.  If these cosmic rays
are not too severely deflected by magnetic fields while in transit,
then their arrival directions will provide crucial information about
their origins.

The High Resolution Fly's Eye (HiRes) experiment, consisting of two
air fluorescence detector sites, has been making
stereo observations of ultrahigh energy cosmic rays since 1999.  The
angular resolution ($\sim 0.6^{\circ}$) achieved by this 
experiment sets the stage for the most precise measurements yet
of small-angle correlation among the highest energy cosmic rays.
In this paper we analyze HiRes events with
energies above $10^{19}$\,eV 
observed between 1999 December and 2004 January.  More details
about this data set are given in Ref.~\refcite{Abbasi:2004ib}.
We also analyze a combination of events from HiRes and 
from the Akeno Giant Air Shower Array (AGASA) with energies above 
$4\times 10^{19}$\,eV.

\section{Analysis}

A standard tool for studying anisotropy is the angular two-point correlation
function.  We define the estimator for the correlation function as
$w(\theta)=N(\theta) / \left<{N}_{mc}(\theta)\right> - 1$, where
$N(\theta)$ is the number of pairs of events in the data sample 
with angular separation
less than $\theta$, and $\left<{N}_{mc}(\theta)\right>$ is the
average number of such pairs in simulated isotropic sets
with the same number of events and same detector acceptance 
in right ascension and declination as the data sample.
Note that this definition of $w(\theta)$ is cumulative over angles
up to $\theta$.  It reveals how
the correlation signal varies as a function of 
the angular threshold for defining a pair.

Fig. 1 shows the angular correlation for the HiRes stereo
events with energies above $10^{19}$~eV, along with the (Poisson) errors
$\sqrt{N} / \left<{N_{mc}}\right>$ (in general, the
variance of correlation estimates may be larger than the 
Poisson variance).  
No significant excess of pairs is observed for any angular scale.

In 1999 and 2000 (Refs.~\refcite{Takeda:1999sg},~\refcite{Hayashida:2000zr})
the AGASA collaboration reported significant clustering
among events above $4\times 10^{19}$\,eV.  In a set of 57 events, 
seven pairs were observed with angular separation less than
$2.5^{\circ}$, against an expected $\left<N_{mc}\right> = 1.5$ pairs, yielding
$w(2.5^{\circ})=3.7$. The probability for 
seven or more pairs evaluated using Monte Carlo is $\sim 0.1\%$.
However, as has been noted in Ref.~\refcite{Finley:2003ur}, this probability
is not indicative of the statistical significance, because of bias
introduced by the {\it a\,posteriori} choice of the energy threshold and 
angular separation.

The angular correlation
estimate can be improved by combining the AGASA and HiRes data.
By itself,
the set of 27 HiRes events above $4\times 10^{19}$\,eV has no pairs
separated by less than $5^{\circ}$, yielding $w=-1$, but with 
large uncertainty.  However, the combination of the two sets 
substantially increases the statistical power over either one individually,
because of the cross-correlation power of the two sets.

To determine the expected number of pairs $\left<N_{mc}\right>$, we simulate
combined sets with 57 events generated using the AGASA acceptance and 
27 generated using the HiRes acceptance.  Although these acceptances are
not identical, they have a large overlap; the resulting value of 
$\left<N_{mc}\right>$ is in fact roughly comparable 
whether one uses all AGASA events, all HiRes events, or a combination.

The two detectors also have different angular resolutions.
While this does not affect simulated isotropic sets,
it could affect the correlation estimate for a real clustering signal: 
a clustering signal of higher significance 
could appear at smaller angular scales because
the HiRes angular resolution is several times sharper
than that of AGASA.  However, since $w(\theta)$ includes all pairs with
separations less than $\theta$, this effect will not lead to a reduced signal
at larger angles.  Therefore rather than attempt to estimate the 
optimal scale for a clustering signal, for the sake of comparison we will
continue to evaluate $w(2.5^{\circ})$ for the combined data set.

Figure 2 shows the results for the 57 AGASA events alone, 
and for the 84 AGASA and HiRes events evaluated jointly.  The addition of the
HiRes data brings one new pair with an AGASA event within $2.5^{\circ}$, 
yielding $N=8$, $N_{mc}=3.0$, and  $w(2.5^{\circ})=1.7$.  
The fraction of simulated sets with eight or more pairs 
is $\sim 1\%$, but it must be emphaiszed that this does not represent a chance
probability because it includes the same bias in the AGASA data 
set noted above.\footnote{A similar but unbiased analysis 
using only the AGASA events observed after the choice
of cuts (i.e. 27 AGASA and 27 HiRes events), 
results in $N=2$, $N_{mc}=1.2$ and $w(2.5^{\circ})=0.7$,
for a chance probability of $34\%$.  Details will be presented in
a separate paper.}
On the other hand, an observation of $w(2.5^{\circ})=3.7$, as seen
originally in the AGASA data alone, 
would have meant the observation of 14 pairs in the combined
data set, corresponding to a $10^{-5}$ deviation from isotropy.

\section{Conclusions}

Angular correlation estimates for the HiRes events above $10^{19}$\,eV reveal
no significant deviations from isotropy at any small angle.
Combining HiRes and AGASA events above $4\times 10^{19}$\,eV improves the
statistics above this threshold relative to the AGASA data set alone,
and results in substantially reduced angular correlation.

\section*{Acknowledgments}

HiRes is supported by the National Science Foundation under
contract numbers NSF-PHY-9321949, NSF-PHY-9322298, NSF-PHY-9974537,
NSF-PHY-0098826, NSF-PHY-0245428, by the Department
of Energy Grant FG03-92ER40732, and by the Australian Research Council.
The cooperation of Colonels E. Fisher and G. Harter, the US Army and
Dugway Proving Ground staff is appreciated.

\begin{figure}\centerline{\psfig{file=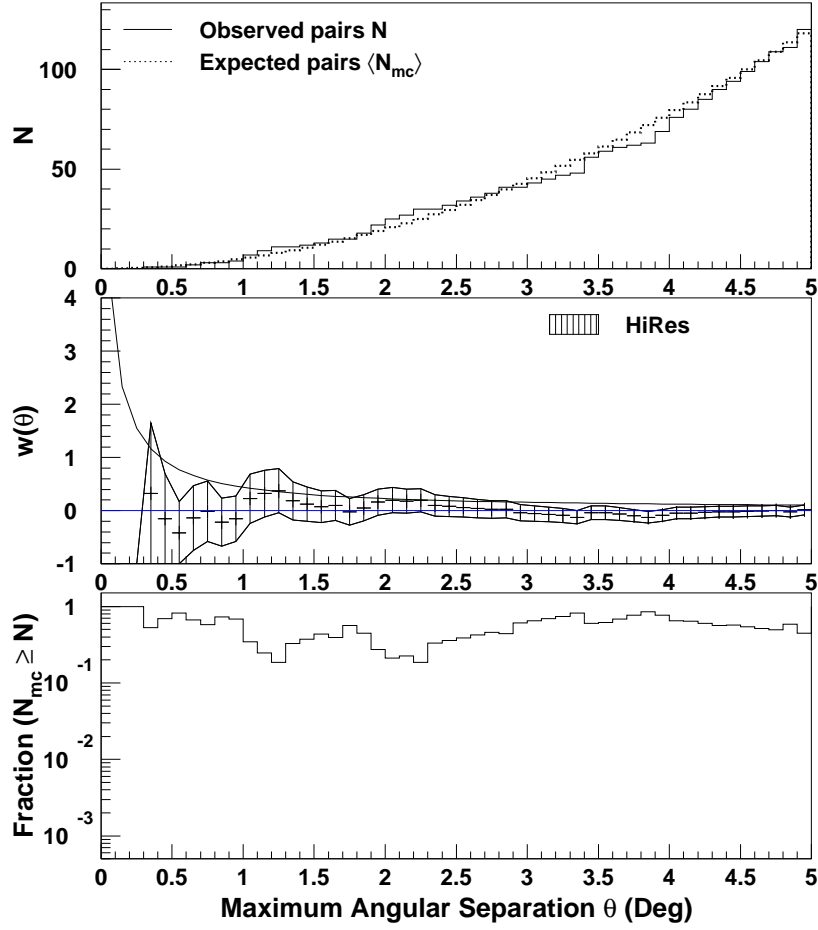,width=11cm}}
\vspace*{8pt}
\caption{the HiRes data set above $10^{19}$\,eV.
Top: Number of observed ($N$) and expected 
($\left<N_{mc}\right>$) pairs of events
as a function of maximum separation angle $\theta$.  Middle: angular 
correlation $w$ and associated uncertainty (see text) for an isotropic 
set (solid curve) and for the data set (vertical bars).
Bottom: the fraction of
simulated Monte Carlo sets with $N_{mc} \ge N$.
Note: values of $w$ in neighboring bins are correlated.}
\end{figure}

\begin{figure}\centerline{\psfig{file=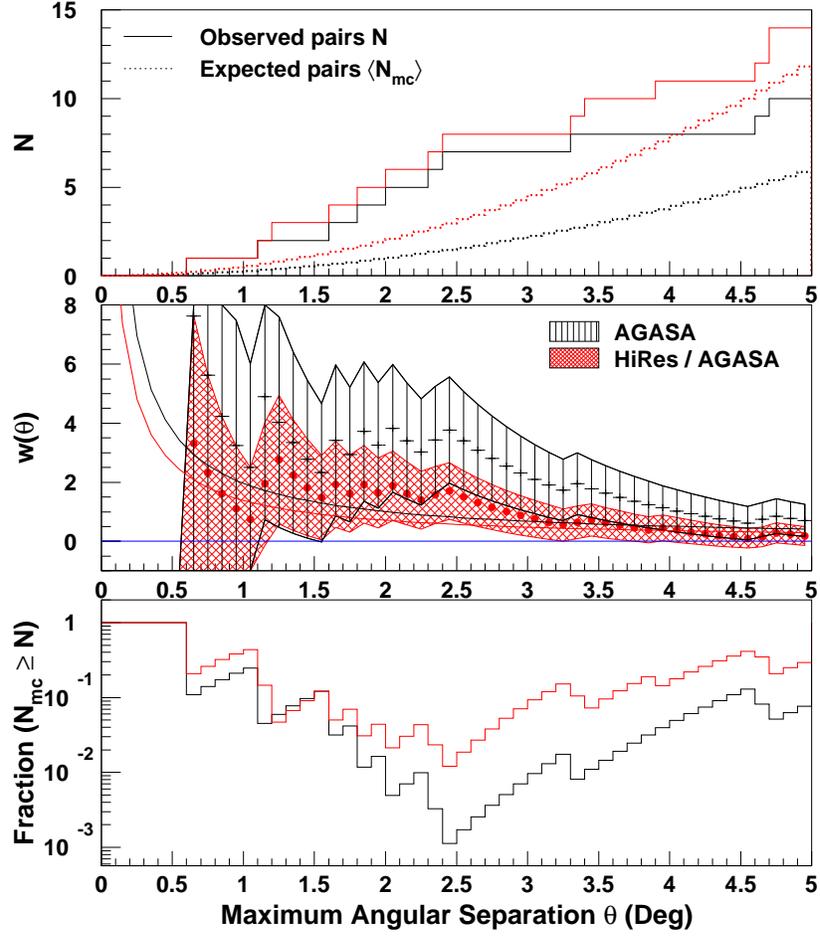,width=11cm}}
\vspace*{8pt}
\caption{the AGASA and the combined 
HiRes/AGASA data sets above $4\times 10^{19}$\,eV. 
Top: Number of observed ($N$) and expected 
($\left<N_{mc}\right>$) pairs of events
as a function of maximum separation angle $\theta$.  Middle: angular 
correlation $w$ and associated uncertainty (see text) for an isotropic 
set (solid curve) and for the data set (vertical bars).
Bottom: the fraction of
simulated Monte Carlo sets with $N_{mc} \ge N$.
Note: values of $w$ in neighboring bins are correlated.}
\end{figure}

\end{document}